\documentstyle[12pt,epsf]{article}
\begin{document}

\catcode`@=11
\long\def\@caption#1[#2]#3{\par\addcontentsline{\csname
  ext@#1\endcsname}{#1}{\protect\numberline{\csname
  the#1\endcsname}{\ignorespaces #2}}\begingroup
    \small
    \@parboxrestore
    \@makecaption{\csname fnum@#1\endcsname}{\ignorespaces #3}\par
  \endgroup}
\catcode`@=12
\newcommand{\newc}{\newcommand}
\newc{\gsim}{\lower.7ex\hbox{$\;\stackrel{\textstyle>}{\sim}\;$}}
\newc{\lsim}{\lower.7ex\hbox{$\;\stackrel{\textstyle<}{\sim}\;$}}
\newc{\gev}{\,{\rm GeV}}
\newc{\mev}{\,{\rm MeV}}
\newc{\ev}{\,{\rm eV}}
\newc{\kev}{\,{\rm keV}}
\newc{\tev}{\,{\rm TeV}}
\newc{\mz}{m_Z}
\newc{\mpl}{M_{Pl}}
\newc\order{{\cal O}}
\newc\CO{\order}
\newc\CL{{\cal L}}
\newc\CY{{\cal Y}}
\newc\CH{{\cal H}}
\newc\CM{{\cal M}}
\newc\CF{{\cal F}}
\newc\CD{{\cal D}}
\newc\CN{{\cal N}}
\newc{\eps}{\epsilon}
\newc{\re}{\mbox{Re}\,}
\newc{\im}{\mbox{Im}\,}
\newc{\invpb}{\,\mbox{pb}^{-1}}
\newc{\invfb}{\,\mbox{fb}^{-1}}
\newc{\yddiag}{{\bf D}}
\newc{\yddiagd}{{\bf D^\dagger}}
\newc{\yudiag}{{\bf U}}
\newc{\yudiagd}{{\bf U^\dagger}}
\newc{\yd}{{\bf Y_D}}
\newc{\ydd}{{\bf Y_D^\dagger}}
\newc{\yu}{{\bf Y_U}}
\newc{\yud}{{\bf Y_U^\dagger}}
\newc{\ckm}{{\bf V}}
\newc{\ckmd}{{\bf V^\dagger}}
\newc{\ckmz}{{\bf V^0}}
\newc{\ckmzd}{{\bf V^{0\dagger}}}
\newc{\X}{{\bf X}}
\newc{\bbbar}{B^0-\bar B^0}
\def\bra#1{\left\langle #1 \right|}
\def\ket#1{\left| #1 \right\rangle}
\newc{\sgn}{\mbox{sgn}\,}
\newc{\br}{\mbox{Br}\,}
\newc{\higgsino}{\widetilde{H}}
\newc{\gluino}{\widetilde{g}}
\newc{\squark}{\widetilde{q}}
\newc{\mumax}{\mu_{\rm max}}
\newc{\wt}{\widetilde}
%
%
\def\NPB#1#2#3{Nucl. Phys. {\bf B#1} (19#2) #3}
\def\PLB#1#2#3{Phys. Lett. {\bf B#1} (19#2) #3}
\def\PLBold#1#2#3{Phys. Lett. {\bf#1B} (19#2) #3}
\def\PRD#1#2#3{Phys. Rev. {\bf D#1} (19#2) #3}
\def\PRL#1#2#3{Phys. Rev. Lett. {\bf#1} (19#2) #3}
\def\PRT#1#2#3{Phys. Rep. {\bf#1} (19#2) #3}
\def\ARAA#1#2#3{Ann. Rev. Astron. Astrophys. {\bf#1} (19#2) #3}
\def\ARNP#1#2#3{Ann. Rev. Nucl. Part. Sci. {\bf#1} (19#2) #3}
\def\MPL#1#2#3{Mod. Phys. Lett. {\bf #1} (19#2) #3}
\def\ZPC#1#2#3{Zeit. f\"ur Physik {\bf C#1} (19#2) #3}
\def\APJ#1#2#3{Ap. J. {\bf #1} (19#2) #3}
\def\AP#1#2#3{{Ann. Phys. } {\bf #1} (19#2) #3}
\def\RMP#1#2#3{{Rev. Mod. Phys. } {\bf #1} (19#2) #3}
\def\CMP#1#2#3{{Comm. Math. Phys. } {\bf #1} (19#2) #3}
\relax
%
%
%
\def\beq{\begin{equation}}
\def\eeq{\end{equation}}
\def\bea{\begin{eqnarray}}
\def\eea{\end{eqnarray}}
%
%
%
\newc{\ie}{{\it i.e.}}          \newc{\etal}{{\it et al.}}
\newc{\eg}{{\it e.g.}}          \newc{\etc}{{\it etc.}}
\newc{\cf}{{\it c.f.}}
%
%
%
%
\def\slash#1{\rlap{$#1$}/} 
\def\Dsl{\,\raise.15ex\hbox{/}\mkern-13.5mu D} 
\def\delsl{\raise.15ex\hbox{/}\kern-.57em\partial}
\def\Ksl{\hbox{/\kern-.6000em\rm K}}
\def\Asl{\hbox{/\kern-.6500em \rm A}}
\def\Qsl{\hbox{/\kern-.6000em\rm Q}}
\def\gradsl{\hbox{/\kern-.6500em$\nabla$}}
%
%
%
\def\bar#1{\overline{#1}}
\def\vev#1{\left\langle #1 \right\rangle}
%

\begin{titlepage}
\begin{flushright}
{MCTP-03-42\\
hep-ph/0310042\\
September 2003\\
}
\end{flushright}
\vskip 2cm
\begin{center}
{\large\bf $B_s\to\mu\mu$ as a Probe of $\tan\beta$ at the Tevatron}
\vskip 1cm
{\normalsize
G.L.~Kane$\,{}^1$, Christopher Kolda$\,{}^{2}$
and Jason E.~Lennon$\,{}^{2}$}\\
\vskip 0.5cm
{\it ${}^1\,$Michigan Center for Theoretical Physics, University of Michigan,
Ann Arbor, MI~~48109, USA\\[0.1truecm]
${}^2\,$Department of Physics, University of Notre Dame,
Notre Dame, IN~~46556, USA\\  
}

\end{center}
\vskip .5cm
\begin{abstract}
Recently it has been understood that flavor-changing
processes mediated by Higgs bosons could be a new and powerful tool
for discovering supersymmetry. In this paper we show that
they may also provide an important method for constraining the parameters
of the minimal supersymmetric standard model (MSSM). 
Specifically, we show that observation of $B_s\to\mu^+\mu^-$ at the Tevatron
implies a significant, model-independent lower bound on $\tan\beta$ in the
MSSM. This is very important because $\tan\beta$ enters crucially in
predictions and interpretations of the MSSM, though it is difficult to
measure. Within specific models, or with other data, the bound becomes
significantly stronger.
\end{abstract}
\end{titlepage}
\setcounter{footnote}{0}
\setcounter{page}{1}
\setcounter{section}{0}
\setcounter{subsection}{0}
\setcounter{subsubsection}{0}


Over the next several years, the Tevatron at Fermilab will be
searching for signs of physics beyond the Standard Model (SM), by
directly producing and observing new particles, and by observing rare
processes at rates inconsistent with SM predictions. In the latter
category falls the search for the rare flavor-changing neutral current
(FCNC) decay $B^0_s\to\mu^+\mu^-$. Observation of this decay at the
Tevatron would necessarily imply new physics since the
predicted rate for this process in the SM is far below the search
capabilities of the machine. However, supersymmetry (SUSY) naturally
predicts large enhancements in the decay rate
mediated by neutral Higgs bosons~\cite{bk,fcnc1,fcnc2,dedes}, and 
in some
cases yields branching ratios three orders of magnitude above the SM. So
observation of the decay would be strong, albeit indirect, evidence in
favor of SUSY.

But observation of $B_s\to\mu\mu$ actually tells us more. We will
show that it is possible to deduce bounds on the
fundamental SUSY parameter $\tan\beta$, 
the ratio of the two Higgs expectation values, from a signal. This is
particularly 
important because $\tan\beta$ is very difficult to measure --- there is
no general technique for measuring it at hadron colliders --- yet
almost all SUSY observables depend on it. And a lower limit may be almost
as useful as an actual measurement because many SUSY observables 
quickly saturate as $\tan\beta$ increases. But
we can obtain a limit precisely because the rate for $B_s\to\mu\mu$
does not saturate. Instead it is
strongly dependent on $\tan\beta$, rising (and falling) as
$\tan^6\beta$. As a secondary feature, we will also be able to obtain
some information on the mass scale of the new SUSY Higgs bosons, since
the branching ratio scales as the fourth power of their masses.

We will begin by considering two very general scenarios in which all
flavor mixing does/does not come from the CKM matrix and in each case
we will arrive at a very strong bound. But we will also consider
some specific models of SUSY breaking and show that even more stringent
bounds can be obtained for these. Our focus will remain on the
Tevatron because it has especially good sensitivity to the signal even
if it does not reach its full luminosity potential. However the
B-factories and LHC can also use this, and related processes, to
further probe $\tan\beta$ and the Higgs sector. Unfortunately, if a
signal is not seen, non-observation 
cannot be used to draw absolute conclusions
about the parameter space of the MSSM other than to rule out specific
model points. It is always possible to choose SUSY parameters in such
a way to push the signal down to the level of the Standard Model where
it would not be observed.

\section{Higgs-Mediated $B\to\mu\mu$}

First, a little theoretical 
background. The question of flavor-changing neutral
currents (FCNCs) mediated by Higgs bosons was first addressed two
decades ago by Glashow and Weinberg~\cite{gw}. It had always been obvious
that the Higgs boson of the {\it minimal}\/ standard model
could not have flavor-violating couplings since the couplings of
fermions to the Higgs field {\it defines}\/ the fermion mass eigenstates 
and thus also defines our notion of flavor. But in models with two
(or more) Higgs fields, the fermion mass eigenbasis can be different
from the Higgs interaction eigenbasis and thus Higgs-mediated FCNCs
can occur. In order to avoid large FCNCs inconsistent with experiment,
the authors of Ref.~\cite{gw} proposed several solutions.
One of those solutions is the
``type~II'' two-Higgs-doublet model: one Higgs field ($H_u$)
couples only to up-type quarks, while the other ($H_d$) couples only
to the down-type: ${\cal L}= \bar Q_L Y_U U_R H_u + \bar Q_L Y_D D_R H_d$.
This guarantees the alignment of the Higgs boson
interaction eigenbasis with the fermion mass eigenbasis. Such a
structure can be protected from quantum corrections by any number of
discrete symmetries under which the two Higgs fields transform differently.

The MSSM possesses the structure of a type-II model classically. But
it possesses no discrete symmetry which can protect this
structure; all such symmetries are broken by the $\mu$-term in the 
superpotential, which must be present 
to avoid disagreement with experiment. 
And though the type-II structure is also protected by
holomorphy of the superpotential, this too is ineffective after SUSY
is broken. Thus new terms are generated in the low-energy Lagrangian
of the MSSM of the form $\bar Q_L \widetilde{Y}_U U_R H_d^\dagger
+\bar Q_L \widetilde{Y}_D D_R H_d^\dagger$. 
Whether such terms will lead
immediately to FCNCs depends on the structure of the new
$\widetilde{Y}_{U,D}$ Yukawa matrices~\cite{bk,fcnc1,fcnc2,dedes}. 

In Ref.~\cite{bk} it was shown that
there are contributions to $\widetilde{Y}_D$ which {\it are}\/
flavor-violating and can have important effects at large $\tan\beta$.
It is actually quite easy to see why. Consider the diagram in
Fig.~1(a) in which charged higgsinos propagate inside
the loop. If we work in a basis in which the down quarks couple
diagonally to $\widetilde H_d^\pm$, then the up quarks have off-diagonal
couplings to $\widetilde H_u^\pm$ proportional to CKM elements. In particular, the
coupling of $\widetilde H_d^\pm$ to $\widetilde b_L b_R$ is just the
bottom Yukawa coupling, $y_b$. But the coupling of 
$\widetilde H_u^\pm$ to $\bar s_L t_R$ is then $y_t V_{ts}$ where $y_t$ is
the top-quark Yukawa coupling. Thus the diagram generates a new
interaction $\bar s_L b_R H_u^\dagger$ with coefficient proportional
to $y_b y_t V_{ts}$. We can rewrite this coefficient as simply $y_b
\epsilon$, where $\epsilon$ includes not only $y_t V_{ts}$ but also
the loop kinematic and suppression factors. (We ignore the phases
induced by the sparticles in the loops; work including them is under
way~\cite{hidaka}.)
\begin{figure}
\centering
\epsfysize=1.25in
\hspace*{0in}
\epsffile{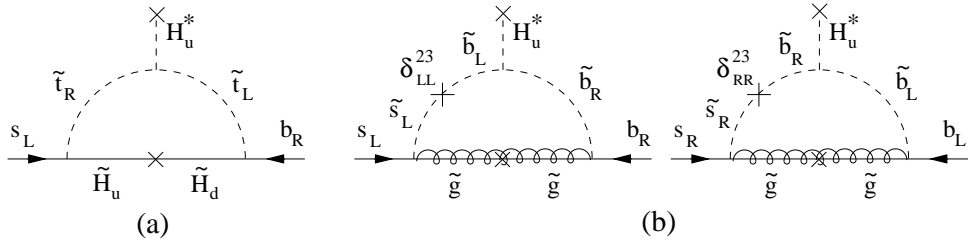}
\caption{New contributions to the $d$-type quark masses in a minimal
  flavor model [(a)] and in general [(a) and (b)].}
\label{fdiags}
\end{figure}

When $H_u$ gets a vev ($v_u$), it contributes an off-diagonal piece 
to the fermion mass matrix:
\beq
{\cal L} = \left(\begin{array}{cc} \bar s_R & \bar b_R \end{array}\right)\,
\left(\begin{array}{cc} m_s & 0 \\ y_b \epsilon v_u
& m_b \end{array} \right) \,\left(\begin{array}{c} s_L \\ b_L
\end{array} \right).
\eeq
This Lagrangian is written in the Higgs interaction eigenbasis of the
fermions, which at tree-level is also the mass eigenbasis; we drop the
first generation for simplicity.
The mass matrix can be diagonalized
by a biunitary transformation which mixes the $s_L$ and $b_L$ 
by an angle $\sin\theta \simeq y_b\epsilon v_u/m_b$. But since
$m_b=y_b \langle H_d \rangle = y_b v_d$, we have 
$\sin\theta\simeq\epsilon \tan\beta$. So although
$\epsilon$ is one-loop suppressed, the factor of $\tan\beta$ can allow
$\sin\theta$ to be ${\cal O}(1)$.

At tree-level, a $\bar bb\to\,$Higgs transition is possible, with the
Higgs decaying to leptons, as in Fig.~2. This occurs by exchange of
$H_d$. However the coupling of the $d$-quark sector to $H_u$ has
shifted the $d$-quark interaction eigenstates away from their 
mass eigenstates. In order to replace the interaction
eigenstates on the external legs with mass eigenstates, we must
replace
\beq b_L \to b'_L=\cos\theta\, b_L + \sin\theta\, s_L\eeq
which induces a $\bar b_R s_L \to \mu\mu$ transition through an $H_d$
Higgs, Fig.~2.
Then the flavor-changing amplitude $\bar bs\to\mu\mu$ is related to
the flavor conserving amplitude $\bar bb\to\mu\mu$ by
\beq
{\cal A}^{\bar bs\to\mu\mu} \simeq \sin\theta {\cal A}^{\bar bb\to\mu\mu}.
\eeq
There are several noteworthy properties of the
Higgs-mediated FCNCs. First, the amplitude for $B_s\to\mu\mu$ scales
as $\tan^3\beta$ at large $\tan\beta$. One factor of $\tan\beta$ come
from $\sin\theta$, the other two come from the $b$- and $\mu$-Yukawa
couplings which scale as $1/\cos\beta$. Thus the branching ratio
scales as $\tan^6\beta$ and provides an incredibly powerful tool for
constraining $\tan\beta$. Second, at large $\tan\beta$, the $H_d$
Higgs doublet is essentially decoupled from the electroweak symmetry
breaking. It contains the physical states $H^0$, $A^0$ and $H^\pm$
which all have roughly equal masses. But
because Higgs flavor changing must
disappear in a model with only one Higgs doublet, the FCNC branching
ratios must decouple as $m_A^4$. But this
does {\it not}\/ mean that the effects decouple as the SUSY mass scale
increases. Rather, in the limit that all supersymmetric masses are
taken heavy, while $A^0$ (or $H_d$)  remains light,
the rate for Higgs FCNCs approachs a finite constant. 

What diagrams contribute to the off-diagonal Higgs couplings?
First, even if all flavor violation stems from the CKM matrix alone, 
there is the one-loop higgsino
diagram of Fig.~1(a) already considered. Models with only this
CKM-induced flavor violation are known as ``minimal flavor violation''
(MFV) models. Note that this kind of flavor violation has nothing to
do with the ``SUSY flavor problem.'' It is always
present in SUSY because it is generated by the flavor violation in the
CKM matrix and cannot be eliminated simply by making superpartners
heavy or by aligning quark/squark mass matrices.

\begin{figure}
\centering
\epsfxsize=5.25truein
\hspace*{0in}
\epsffile{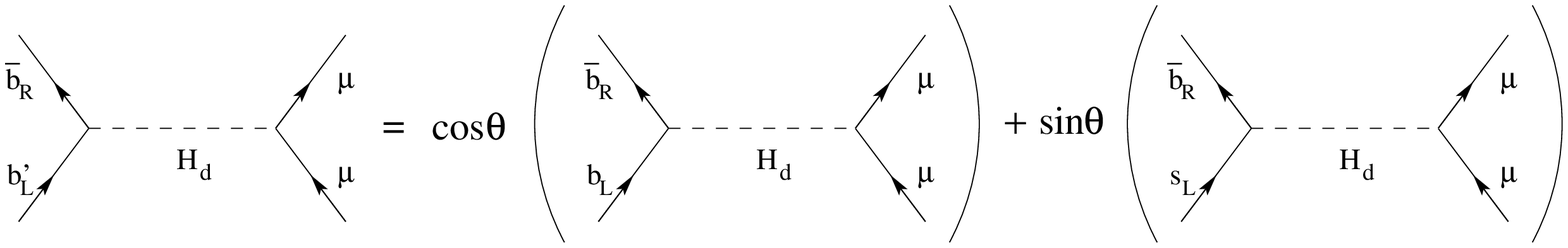}
\caption{Diagram for $\bar bb\to\,$Higgs$\,\to\mu\mu$ in the
interaction (primed) 
eigenbasis becomes $\bar bs\to\,$Higgs$\,\to\mu\mu$ in the mass
eigenbasis, suppressed by $\sin\theta$, the $s_L$--$b_L$ mixing angle.}
\label{decay}
\end{figure}

More generally
there are also gluino loop diagrams contributing to the same
interactions (Figure~1(b)), 
if flavor-violating $LL$ or $RR$ squark mass insertions are
non-zero. Such an insertion can be generated in minimal models by
renormalization group running of the squark mass matrices, or may
appear in the underlying theory.
We make no assumptions in our general analysis
about the underlying source of this flavor-changing in the squark
sector, and we refer to this as general flavor violation.

Because this new source of FCNCs emerges from the Higgs sector, it
preferentially generates processes involving heavy fermions.
Thus the channels in which
Higgs-mediated FCNCs can most easily be observed are
$B_{s,d}\to\tau\tau$ and $B_{s,d}\to\mu\mu$. The channels with final
state muons are suppressed relative to those with final state taus
by $(m_\mu/m_\tau)^2$ but are much cleaner experimentally. The
channels with initial state $B_d$ are suppressed relative to $B_s$ by
$(V_{td}/V_{ts})^2$. Thus a machine which produces an abundance of
$B_s$ and can cleanly tag and measure muons is ideal for studying this
physics. This machine is precisely the Tevatron, either in CDF or
D0. The current limits on $\br(B_{s,d}\to\mu\mu)$ are provided by CDF from
Run~I at $2.0\times 10^{-6}$ and $6.8\times 10^{-7}$
respectively. Given enough luminosity, the $B$-factories may
be able to corroborate any Tevatron discovery using $B_d\to\mu\mu$ or, if
a suitable technique is found, $B_d\to\tau\tau$.

If CDF and D0 each record even $2\invfb$ of data during Run~II of the
Tevatron, now underway, they can
probe the $B_s\to\mu\mu$ branching fraction
below the $10^{-7}$ level. (See, \eg, Ref.~\cite{arnowitt} for a more
detailed study of the CDF capabilities.) We will for illustration assume
that the Tevatron can discover a signal in $B\to\mu\mu$ if its
branching ratio is greater than $10^{-7}$. Because the branching
ratios scale as high powers of the input parameters 
($m_A^4$ and $\tan^6\beta$), small changes in the Tevatron
capabilities will generate infinitessimal changes in our results.
And once a signal is discovered, the numerical analysis can be redone
to obtain more precise limits.

We demonstrate below that there is a minimum value of $\tan\beta$
consistent with a signal at the $10^{-7}$ level, and calculate it.
We also
consider the interplay between the pseudoscalar mass and $\tan\beta$,
a result which could be important in searches for the additional Higgs
bosons at future colliders. Both of these analyses will be done in
full generality within the MSSM. We will not constrain ourselves
initially to
any particular class of MSSM models, such as supergravity- or
anomaly-mediated SUSY-breaking. This can be done because there are
only a limited number of parameters on which the branching ratio will
depend and these few parameters can be studied without further
simplifications or assumptions. However in particular classes of models the
constraints on $\tan\beta$ and $m_A$ are stronger and
therefore observation of a signal provides even more information.
We will consider specific SUSY models after doing the general analysis
and we will find bounds on $\tan\beta$ much stronger than in the
general case.

\section{Minimal Flavor Violation: {\sl Higgsino Contribution}}

Regardless of any details of SUSY-breaking, the higgsino contribution
of Fig.~1(a) to Higgs-mediated flavor-changing processes must be
present. First we will consider the case where {\it only}\/ this
contribution is present and then generalize in the next section. Thus
we begin by considering the MFV models.

The flavor-changing 
contribution of the higgsino loop diagram is encoded in a
new dimensionless parameter $\kappa_{\higgsino}$ such that
\beq
\br(B_s\to\mu\mu)\simeq \frac{G_F^2}{8\pi}\, \eta_{\rm QCD}^2 
m_{B_s}^3 f_{B_s}^2 \tau_{B_s} m_b^2
\,m_\mu^2\left(\frac{\tan^2\beta}{\cos^4\beta}
\right) \left(\frac{\kappa_{\higgsino}^2}{m_A^4}\right).
\label{eq:br}
\eeq
Here $\eta_{\rm QCD}\simeq 1.5$ is the QCD correction due to running
between the SUSY and $B_s$ scales, and we take
$f_{B_s}=220\mev$. Note that the fraction $\tan^2\beta/\cos^4\beta$
approaches $\tan^6\beta$ at large $\tan\beta$.

The parameter $\kappa_{\higgsino}$ is calculated to 
be~\cite{bk}
\beq
\kappa_{\higgsino} = -\frac{G_F \,m_t^2 \,V_{ts} V_{tb}}
{4\sqrt{2}\pi^2\sin^2\beta}\,\mu A_t\, f(\mu^2, m_{\widetilde
t_L}^2, m_{\widetilde t_R}^2)
\eeq
where $m_{\widetilde t_L}$ and $m_{\widetilde t_R}$ are the left- and
right-handed top squark masses,
$\mu$ is the superpotential Higgs mass parameter,
$A_t$ is the top-squark trilinear term, and the
function $f$ is defined in Ref.~\cite{bk}. The important thing to know
about $f$ is that it is positive definite, symmetric in its inputs and 
$f(x,y,z)\sim 1/\mbox{max}(x,y,z)$ up to a constant of
${\cal O}(1)$. In particular, $f(m^2,m^2,m^2)=1/(2m^2)$ and
$f(m^2,m^2,0)= 1/m^2$.

In order to maximize the function $\kappa_{\higgsino}$ (and therefore
maximize the branching fraction one can obtain for a given
$\{m_A,\tan\beta\}$) we need only consider the four parameters
$m^2_{\widetilde t_L}$, $m^2_{\widetilde t_R}$, $\mu$ and $A_t$. This is
easiest to do by considering the limits in which $\kappa_{\higgsino}$
could become large.

First
consider the limit in which $\mu$ is much larger than the
squark masses. Then $\kappa_{\higgsino}\sim A_t/\mu$ since $\mu$ appears
both in $f$ and as a prefactor. Thus it appears that
$\kappa_{\higgsino}$ can grow unabated as $A_t$ becomes large. Such a
runaway behavior would generate arbitrarily large branching ratios and
invalidate our claim to a bound on $\tan\beta$. However,
there is a ``cosmological'' limit to $A_t$: as $A_t$ increases far
beyond the squark masses, the usual electroweak vacuum become unstable
with the true vacuum breaking QED and QCD~\cite{colorbreaking}. 
We apply this constraint by applying the famous condition
\beq
A_t^2< 3\,(m^2_{\widetilde t_L}+m^2_{\widetilde t_R}+m_2^2).
\eeq
(The last term, $m_2^2$, is the mass parameter for $H_u$ appearing in the
Higgs potential. Since $m_2^2<0$ is necessary to break the electroweak
symmetry, we can maximize $A_t$ by setting $m_2^2=0$.)
Because the squark masses are much smaller than $\mu$ in this limit,
then $A_t/\mu$ can not become much larger than unity at best.
A similar argument holds for large $A_t$ but with $\mu$ smaller than
the squark masses, except now $\kappa_{\higgsino}$ is more highly
suppressed, $\kappa_{\higgsino}\sim A_t\mu/m_{\widetilde t}^2$.

The actual limit is obtained when $\mu$ becomes large along with one of
the squark masses (with the other as small as possible). Then
$f(\cdots)\to 1/\mu^2$. But the QED/QCD-breaking constraint still
limits $A_t$, though now it becomes $A_t<\sqrt{3}\,\mu$ so that
$\mu A_t f < \sqrt{3}$ which implies that 
\beq
\kappa_{\higgsino}< \frac{0.011}{\sin^2\beta}.
\eeq
Plugging into Eq.~(\ref{eq:br}) immediately gives
\beq
\br(B_s\to\mu\mu)< \frac{5\times 10^{-6}\gev^4}{m_A^4\cos^6\beta}.
\eeq
Thus for $m_A=100\gev$ (approximately its current lower limit)
and a branching ratio greater than $10^{-7}$, we
deduce in an MFV scenario that $\tan\beta>11$. As $m_A$ increases, the
bound on $\tan\beta$ also increases rapidly; for example, if 
$m_A>200\gev$, one must have $\tan\beta > 18$. In general,
\beq
\tan\beta > 11\,\left(\frac{m_A}{100\gev}\right)^{\frac23}
\left[\frac{\br(B_s\to\mu\mu)}{1\times 10^{-7}}\right]^{\frac16}
\label{eq:mfv}
\eeq
where we have taken $1/\cos\beta\simeq\tan\beta$ for
large $\tan\beta$.

\begin{figure}
\centering
\epsfysize=2.2in
\hspace*{0in}
\epsffile{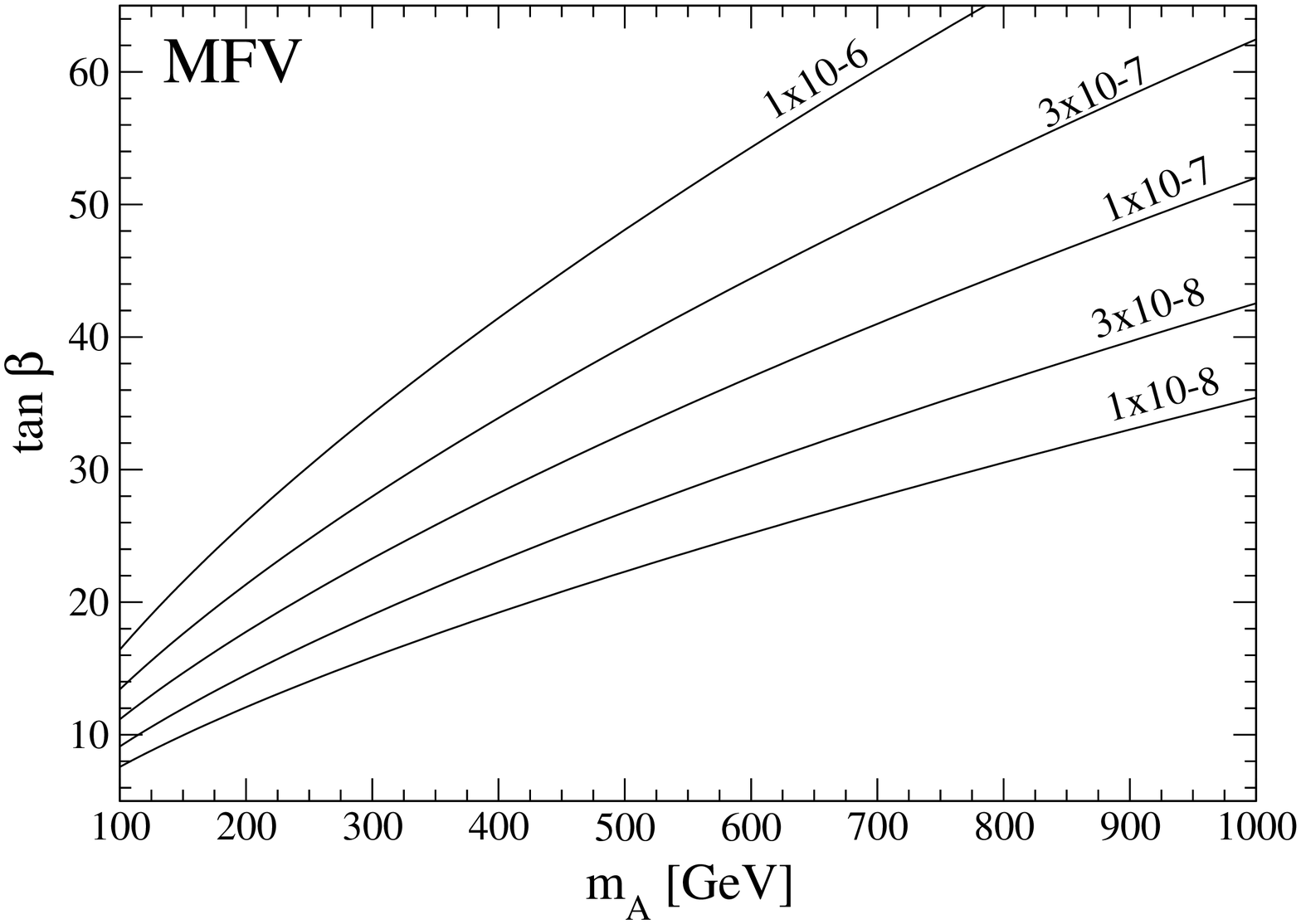}
\caption{Contours of maximum allowed value for $\br(B_s\to\mu\mu)$
(labelled) as
a function of $\{m_A, \tan\beta\}$ for minimal flavor violation. 
For a given branching ratio, 
the allowed region is above the labelled line.}
\label{mfvfig}
\end{figure}

In Figure~\ref{mfvfig}, we plot contours representing the maximal
value of the $B_s\to\mu\mu$ branching ratio consistent with a given choice of
$m_A$ and $\tan\beta$. Given a measured value of the branching ratio,
only the region above the line is consistent in an MFV scenario.

\section{General Flavor Violation: {\sl Gluino Contributions}}

In the case of MFV, only the higgsino loop has any significant effect
in producing flavor violation. However once one moves away from truly
minimal flavor violation, a new source arises.
In a truly general scenario, the
$\gluino$-$\squark$-$q$ couplings can be flavor changing. In the
mass insertion approximation, these violations are moved to
the squark 
propagators where they appear as bilinear interactions which mix flavors
and even chiralities of the squarks. The size of the insertions is
parametrized by a dimensionless quantity, $\delta_{ab}^{ij}=(\Delta
m^2_{ab})_{ij}/m^2_{ab}$ where ${a,b}$ are either $L$ or $R$, and
${i,j}$ label the squark generation. We are only interested in the 
down-sector
$\delta_{LL}^{23}$ and $\delta_{RR}^{23}$ insertions for this work. 
At present there are no strong
experimental bounds on these insertions and so it is possible that the
$\widetilde s_L$ and $\widetilde b_L$ (or $\widetilde s_R$ and
$\widetilde b_R$) are maximimally mixed, or 
$\delta_{LL,RR}^{23}\simeq 1$. Of course, in particular models a large
$\delta_{LL,RR}^{23}$ could lead to large $B\to K^*\gamma$ or
$B\to\phi K_S$, but constraints from these processes are sensitive to
$\delta_{ab}^{23}/m^2_{ab}$ while $B\to\mu\mu$ is sensitive to
$\delta_{ab}^{23}$ alone. Since we allow very large flavor-mixing, we
must take care to note that the mass insertion approximation breaks
down as $\delta$ approaches one; we will discuss this issue shortly.

The new gluino-induced source of flavor changing in general models
generates new diagrams in $B_s\to\mu\mu$ with $\widetilde b$ and 
$\widetilde s$ squarks and gluinos in the loop; see Fig.~1(b).
For our purposes here we can ignore the higgsino contributions
since their maximal size is smaller than that of the gluino loop if
$\delta_{LL,RR}^{23}\sim 1$. We
will also assume that only one of the two diagrams in Fig.~1(b)
dominates the branching ratio; when there is data, a fully combined
analysis should be done. Specifically we will choose the
$\delta_{LL}^{23}$ diagram to dominate, though our results are
identical if the $\delta_{RR}^{23}$ diagram dominates instead.
We will discuss more realistic scenarios at the end of the section.

We parametrize the gluino contribution as
\beq
\kappa_{\gluino} = \frac{2\alpha_s}{3\pi}\, \delta_{LL}^{23}\,
\mu M_3 f_2(m^2_{\widetilde{s}_L},m^2_{\widetilde{b}_L},
m^2_{\widetilde{b_R}},M_3^2)
\label{eq:gluino}
\eeq
where the function $f_2$ is defined in Ref.~\cite{bk2}. The branching ratio
is calculated from $\kappa_{\gluino}$ just as for $\kappa_{\higgsino}$ in
Eq.~(\ref{eq:br}). Notice in particular the lack of a $V_{ts}$
suppression in $\kappa_{\gluino}$; this will allow much smaller values
of $\tan\beta$ to be consistent with an experimental signal.

There are two limits in which we can simplify the above
expression. For small $\wt{b}_L$--$\wt{s}_L$ mixing, we can usually 
take $m^2_{\wt{s}_L}=m^2_{\wt{b}_L}$ in Eq.~(\ref{eq:gluino}) and then
work in the mass insertion approximation.
However, if there is a large hierarchy in between $\wt{s}_L$ and
$\wt{b}_L$, or if there is large mixing, that approximation breaks
down. Instead, it is far easier to work directly in
the basis of the mass eigenstates. Defining $\widetilde{d}_{Li}$
$(i=1,2)$ as $\widetilde{b}_L\cos\theta +\widetilde{s}_L\sin\theta$
and its orthogonal combination, we must replace the product
$\delta_{LL}^{23} f_2(\cdots)$ in Eq.~(\ref{eq:gluino}) with
\beq
\cos\theta\sin\theta\left\lbrace f(m_{\widetilde{d}_{L1}}^2,\cdots)
-f(m_{\widetilde{d}_{L2}}^2,\cdots)\right\rbrace
< \frac12 f(m_{\widetilde{d}_{L1}}^2,\cdots).
\label{eq:sin}
\eeq
The inequality follows only from the positivity of the function
$f$. Given more information in the future about the squarks (masses or
lower bounds) 
it may be possible to find a lower bound on the second term and thus
an improved upper bound on the branching fraction.

Like the MFV case there is a possible runaway behavior in computing
the diagrams in Fig.~1(b), allowing for arbitrarily large branching
ratios. In the MFV case, this arose as $A_t\to\infty$, which we cut
off by demanding that color-breaking minima deeper than the SM minimum
not appear. In a general case, this arises as $\mu\to\infty$ for which
the color-breaking constraint is useless. However there are equally
powerful constraints which rely on fine-tuning arguments, which have
recently been strengthened~\cite{klnw}. In
particular, the $\mu$-parameter appears in the Higgs potential and it
is the minimization of this potential which must supply the weak
scale. There is a well-known relation among the $\mu$-term and other
Higgs soft mass terms which must together generate the scale $m_Z$:
\beq
\frac12 m_Z^2 = \frac{m_{H_d}^2 - m_{H_u}^2\tan^2\beta}{\tan^2\beta
-1} -\mu^2.\label{rewsb}
\eeq
If $|\mu|\gg m_Z$ then a fine-tuning must be arranged among $m_{H_u}$,
$m_{H_d}$ and $\mu$ in order to generate $m_Z$ on the left hand side
of the Eq.~(\ref{rewsb}). We will require 
$|\mu|<500\gev$, which is a statement that we allow less
fine-tuning in the electroweak potential than about one part in 60. 
However
in our expressions we will show how to scale our results for other
choices of $\mu$, in case the reader wishes to apply their own
fine-tuning constraint.

\begin{figure}
\centering
\epsfysize=2.5in
\hspace*{0in}
\epsffile{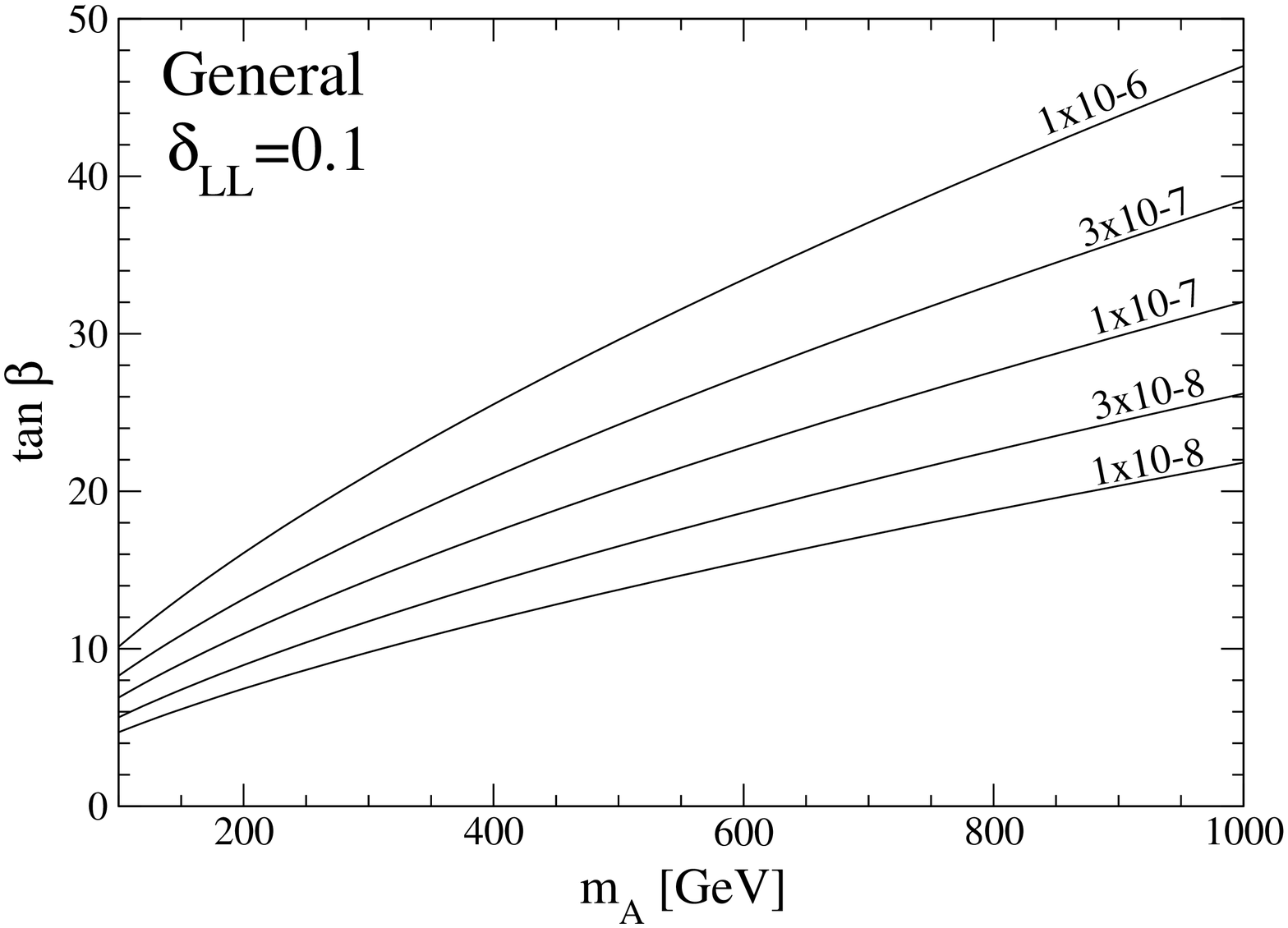}
\caption{Contours of maximum allowed value for $\br(B_s\to\mu\mu)$
(labelled) as
a function of $\{m_A, \tan\beta\}$ for non-minimal flavor
violation with $\delta_{LL}^{23}=0.1$. For a given branching ratio, 
the allowed region is above the labelled line.}
\label{nmfvfig}
\end{figure}

In order to calculate the upper bound on the branching ratio, it is
useful to note that the product $M_3 f(M_3^2,m^2_{\widetilde{q}_L},
m^2_{\widetilde{q}_R})$ reaches its upper bound when the squarks are
as light as possible and degenerate, and $M_3\simeq 2.1 m_{\widetilde q}$.
Then one can derive the semi-analytic bound (assuming maximal mixing):
\beq
\kappa_{\gluino} < 0.009\, 
\left|\frac{\mu}{500\gev}\right|
\eeq
leading to 
\beq
\br(B_s\to\mu\mu)<\frac{3.0\times10^{-3}\gev^4}{m_A^4}
\left|\frac{\mu}{500\gev}\right|^2\,\tan^6\beta .
\eeq
and
\beq
\tan\beta > 4\, 
\left(\frac{m_A}{100\gev}\right)^{\frac23}
\left[\frac{\br(B_s\to\mu\mu)}{1\times 10^{-7}}\right]^{\frac16}
\left|\frac{500\gev}{\mu}\right|^{\frac13}.
\label{eq:nmfvmax}
\eeq
Contours of the maximum allowed branching ratio are plotted in
Fig.~\ref{nmfvfig} as a function of $m_A$ and $\tan\beta$ for the case
of maximal mixing.
For small mixing we can return to the mass insertion approximation in
which case
\beq
\tan\beta > 7\, 
\left(\frac{0.1}{\delta_{LL}^{23}}\right)^{\frac13}
\left(\frac{m_A}{100\gev}\right)^{\frac23}
\left[\frac{\br(B_s\to\mu\mu)}{1\times 10^{-7}}\right]^{\frac16}
\left|\frac{500\gev}{\mu}\right|^{\frac13}.
\label{eq:nmfvmin}
\eeq

The bounds presented in this section should be considered the absolute
limits on $\tan\beta$ given a signal at the Tevatron at the $10^{-7}$
level. Nowhere did we inject theoretical constraints stemming from
some particular model of SUSY. On the other hand, if we had a model,
or other information, we could devise much stronger bounds. For
example, many observables in SUSY depend on $\mu$, so that once SUSY
is found, we will very quickly have constraints on $\mu$, which would
allow more precise bounds. And if we have some theoretical or
experimental reason to believe in a particular model of SUSY-breaking,
then we know even more.
As we will discuss in Section~\ref{models}, models
produce bounds which are much stronger than the general case.

\subsection{``Nearly-Minimal'' Flavor Violation}

For most purposes, the bounds in the previous section are too
general. SUSY models which are truly generic and general have an
embarassing wealth of flavor-changing mass insertions like 
$\delta_{LL}^{23}$. Most of these insertions must be miniscule
in order to avoid dangerously large FCNCs in well-studied processes,
such as $K^0$-$\bar K^0$ mixing. Thus one usually imposes on SUSY some
form of organizing principle which prevents most of these insertions
from becoming large, whether that be mass degeneracies or mass matrix
alignment. 

However even in models which naturally solve the SUSY flavor problem,
one usually still expects some flavor violation to reappear by way of
renormalization effects. The most common source of these is the
presence of the up-quarks Yukawa matrix in the renormalization group
equations for the left-handed squarks. Because all Yukawa matrices
cannot be simultaneously diagonalized, non-zero $\delta_{LL}^{ij}$
will be generated. (The $RR$ mass insertions will not be generated at
leading order.) However there is a significant difference between
these mass insertions and those considered in the general case above:
here $\delta_{LL}^{ij}$ will be proportional to the corresponding
element of the CKM matrix. For example, $\delta_{LL}^{23}$ will be
naturally suppressed by $V_{ts}$. Thus one typically finds 
$\delta_{LL}^{23}\lsim {\cal O}(V_{ts})$ which in turn would force
$\tan\beta\gsim 9$ for $\br(B_s\to\mu\mu)>10^{-7}$. However this bound
is not a true limit -- the model can be tuned in order to obtain the
same branching ratio but with much smaller values of $\tan\beta$.

Thus, while one expects from theoretical arguments that $\tan\beta$
must be greater than 9 in order to see a signal at the Tevatron, the
true mathematical limit remains at $\tan\beta=4$ as found in the previous
section, until the origins of SUSY flavor physics are better understood.

\subsection{Relation to Other Observables}

The bound on $\tan\beta$ in the general case depends strongly on the
$LL$ (or $RR$) mixing present in the strange--bottom squark sector. 
But there are
several other key observables which also depend on this mixing and
which should be measured or constrained over the next few years. Key
among them are $\bar B_s$--$B_s$ mixing, parametrized by $\Delta M_s$,
and the rate and CP asymmetry in $B_d\to\phi K_S$. On the subject of
$B_d\to\phi K_S$, we will not say very much, but refer the reader to
Ref.~\cite{kkkpww} for a full discussion.

The situation for $\Delta M_s$ is more interesting. In the presence of
a single, dominating $LL$ (or $RR$) mass insertion, a calculation of
$\Delta M_s$  yields the approximate formula:
\beq
 \Delta M_s \simeq \left|\left(\Delta M_s\right)_{\rm SM} +
\left(\frac{3500\gev}{\widetilde m}\right)^2
\left(\delta_{LL}^{23}\right)^2\right|
\eeq
where $(\Delta M_s)_{\rm SM}$ is roughly 15 to
$20\,\mbox{ps}^{-1}$. The same formula also holds with the
substitution $LL\to RR$.
(To obtain this
simple result, we have followed the calculations of Ref.~\cite{bbbar}
for $B_d$--$\bar B_d$ mixing, appropriately modified, and
assumed all squark and gluino masses are
essentially degenerate at $\widetilde m$.) Since $\delta_{LL}^{23}$ can
be complex, the SUSY contributions can either add to or subtract from
the SM prediction for $\Delta M_s$. A measurement of $\Delta M_s$ near
its SM value would indicate either large $\tilde m$ or small
$\delta_{LL}^{23}$, either one of which would suppress $B_s\to\mu\mu$
and thus tighten our bound on $\tan\beta$. 
Once SUSY partners are
discovered, the value of $\widetilde m$ will be known and a very
stringent bound on $\delta_{LL}^{23}$ can be extracted from the
$\Delta M_s$ data, strengthening our bound on $\tan\beta$ further.

\section{Supersymmetric Models} \label{models}

In the previous sections, we performed very general
analyses in determining what could be
learned about the SUSY parameter space by measuring
$\br(B_s\to\mu\mu)$. Now we ask the same question, but in the
context of specific models of SUSY-breaking, namely the three most
commonly studied models: minimal supergravity models (mSUGRA, also
called the constrained MSSM), 
minimal gauge-mediated (GMSB) models, and minimal
anomaly-mediated (AMSB) models. Realistic models will necessarily
yield much stronger bounds on $m_A$ and $\tan\beta$ for two reasons:
first, there is no reason for their contributions to FCNCs to be
maximal, and second there are often correlations with other processes
which suppress the flavor-changing contributions. Several analyses of
$B_s\to\mu\mu$ in the context of SUSY models have been
completed~\cite{models} and a more detailed
study of this will appear shortly~\cite{kl}; here we summarize
some of the important results of this last study. Namely we will find
bounds on $\tan\beta$ in specific models for an observed 
$\br(B_s\to\mu\mu)$ and discuss their implications.

For each of the three models to be discussed, we will vary all the
free parameters of the models and at each ``model point'' calculate
the $B_s\to\mu\mu$ branching fraction~\cite{softsusy}. 
However we will also apply
several phenomenological constraints to avoid points which are
unphysical. The most important constraints come from the muon magnetic
moment~\cite{mdm} 
and the $b\to s\gamma$ transition~\cite{bsgamma}. 
For the former, we will
demand that the SUSY constributions to $(g-2)_\mu$ fall between
$-5\times 10^{-10}$ and $57\times 10^{-10}$.
For the latter, we will demand that calculated rate for $b\to s\gamma$
fall between $2.1\times10^{-4}$ and $4.5\times 10^{-4}$. 
In both cases these are rather wide windows. For the
muon magnetic moment, a wide window is required by the inconsistency
of the $e^+e^-$ and $\tau$ data sets which are used to calculate the
standard model hadronic contributions to $g-2$. For $b\to s\gamma$ we
use a wide window because of the uncertainties in the next-to-leading
order calculation of the SUSY contributions.

\subsection{Minimal supergravity}
 
Minimal supergravity (mSUGRA) actually contributes fairly efficiently to
Higgs-mediated FCNCs. It has three key ingredients that help to generate
large branching ratios. First, mSUGRA models often have large $A_t$
trilinear terms, necessary for the higgsino contribution.
Second, the extended running from the
GUT scale down to the weak scale generates considerable mixing between
the second and third generation squarks, allowing for a sizable gluino
contribution.
Finally, the pseudoscalar Higgs does not come out particularly heavy. 

\begin{figure}
\centering
\epsfysize=3in
\hspace*{0in}
\epsffile{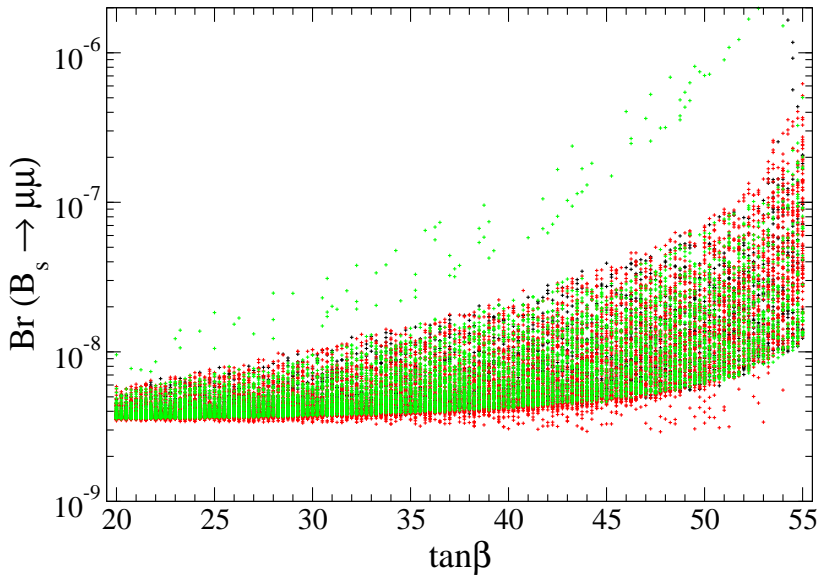}
\caption{Correlation of $\tan\beta$ and $\br(B_s\to\mu\mu)$ in a
representative sample of mSUGRA models.}
\label{sugrafig}
\end{figure}

In Fig.~\ref{sugrafig}(a), we show a scatter plot of points in the
mSUGRA model space, each one representing a distinct choice of 
$\tan\beta$, $M_{1/2}$, $m_0$, and $A_0$ (we take $\mu>0$ only),
consistent with all experimental and theoretical constraints. 
There are several important lessons to draw from the figure. One
should immediately note that mSUGRA models exist, consistent with all
other constraints, that have rates for $B_s\to\mu\mu$ right up to the
CDF bound; in fact, models were found that passed all other
constraints, except the $B_s\to\mu\mu$ bound, indicating that
$B_s\to\mu\mu$ is {\it already}\/ a non-trivial constraint on mSUGRA models. 

We also notice that mSUGRA models with a branching ratio greater than
$10^{-7}$ require $\tan\beta>40$. Thus observation of $B_s\to\mu\mu$
at Run~II would certainly indicate very large $\tan\beta$ if mSUGRA is
the correct model. 

One also notices another interesting feature in the
figure, namely a bifurcation of the points at the largest branching
fractions. The points in the upper arm all have $A_0<0$ while all the
points with $A_0>0$ lie in the main body of points. This behavior is
rooted in the $b\to s\gamma$ constraint. 
For $A_0>0$, models with very large $B_s\to\mu\mu$ also
have chargino contributions to $b\to s\gamma$ which tend to cancel the SM.
Thus the resulting $\br(b\to s\gamma)$ is too small and
we throw the model point out. As the SUSY mass scales increase, the
cancellation goes away, but the rate for $B_s\to\mu\mu$ is also
suppressed. However, the cases with $A_0<0$ exhibit a very different
behavior. For light spectra, these models can have SUSY contributions to
$b\to s\gamma$ much larger than those of the SM, but with opposite
sign, so that after adding the two pieces, the resulting rate is still
roughly consistent with the SM and with experiment. Then as the SUSY
mass scale is increased, the two pieces start to cancel to zero,
ruling out the model points. As the SUSY scale is further increased,
the SUSY contribution decouples and the SM once again dominates. Thus
a gap is generated where the models predict $\br(b\to
s\gamma)\simeq 0$.

Thus models with $A_0>0$ can only be observed in Run~II if
$\tan\beta>50$ while models with $A_0<0$ can have $\tan\beta$ down to 40.
The upshot of this is that a measurement of $B_s\to\mu\mu$ and an
independent measurement of $\tan\beta$ could indicate that
$A_0<0$. However, these measurements alone cannot be used to show that
$A_0>0$. 

\subsection{Anomaly Mediation}

The simplest models with pure anomaly mediation have only one free
parameter, an overall mass scale, but 
are not phenomenologically viable. But one can define a ``minimal''
anomaly-mediated model which has two free parameters: an overall mass
scale for the anomaly mediation, and a separate mass scale just
for the scalars. We considered this minimal model, 
by varying both of these parameters, as well as
$\tan\beta$, and studying the resulting spectra. In principle, AMSB
shares many of the same features that make mSUGRA a nice candidate for
discovery of $B_s\to\mu\mu$. In practice, this is only partially
true. In AMSB the $b\to s\gamma$ constraint is much stronger and rules out
many model points where a large Higgs-mediated FCNC signal would have
been predicted. We do not find any models with observable
branching fractions of $B_s$ into muons $(>10^{-7})$, consistent with
the result of Baek \etal~\cite{models}. However, we note that this result
depends very strongly on the $b\to s\gamma$ calculation; changes in
the NLO calculation could easily allow for much larger (or smaller) 
$B_s\to\mu\mu$ branching fractions.

\begin{figure}
\centering
\epsfysize=3in
\hspace*{0in}
\epsffile{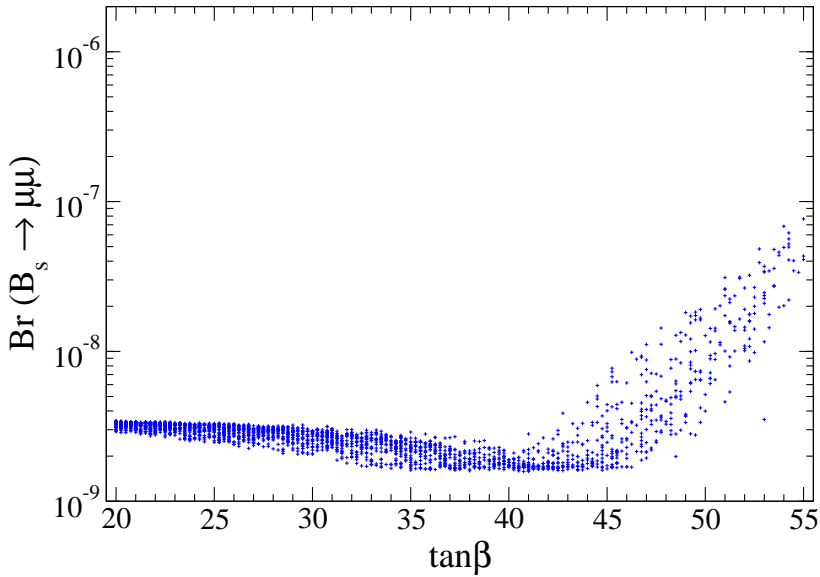}
\caption{Correlation of $\tan\beta$ and $\br(B_s\to\mu\mu)$ in a
representative sample of AMSB models.}
\label{amsbfig}
\end{figure}

A scatter plot of the
allowed parameter space is shown in Fig.~\ref{amsbfig}(a).
This figure again shows a bifurcation into two separate regions, though
this time much more distinct than in the mSUGRA case. Here there is no
fundamental difference between the two regions, but the basic
explanation is similar to the case of the last section. For AMSB
models with large $\tan\beta$ and intermediate masses, the SUSY
contribution to $b\to s\gamma$ tends to cancel the SM, predicting a
rate which is too small. For large masses, the SUSY contribution
decouples, but then so does $B_s\to\mu\mu$. However, there is a region
of very light masses where the SUSY contribution is roughly twice the
SM contribution, but with opposite sign, so that their sum squared is
consistent with data. This last region is the
upper-right region of the figure. There we see that one can have
$\br(B_s\to\mu\mu)$ are large as $10^{-7}$ only if $\tan\beta>55$.

\subsection{Gauge Mediation}

\begin{figure}
\centering
\epsfysize=3in
\hspace*{0in}
\epsffile{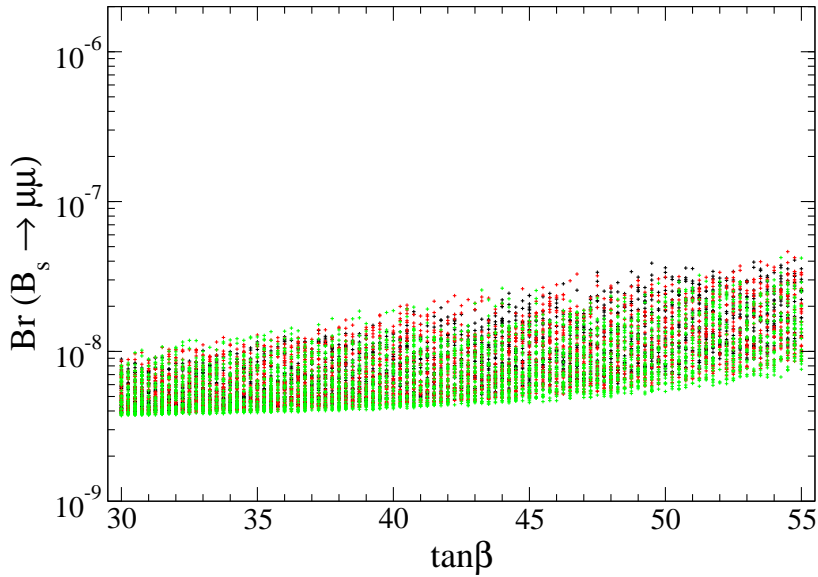}
\caption{Correlation of $\tan\beta$ and $\br(B_s\to\mu\mu)$ in a
representative sample of GMSB models.}
\label{gmsbfig}
\end{figure}

Unlike the previous two cases, one would not expect a large
Higgs-mediated FCNC signal in models with gauge mediation. 
Whereas mSUGRA has a large $A_t$ and lots of running to generate
squark mixing, gauge-mediated models have neither.
At the mediation scale, $A_t=0$ (so no higgsino contribution)
and there is little running unless the mediation scale is large (so no
gluino contribution). Indeed, we find that it is very
difficult to generate observable signals at the Tevatron for GMSB
models, even without invoking constraints from $b\to s\gamma$.
This is best summarized by Fig.~\ref{gmsbfig}, where no
models are found with $\br(B_s\to\mu\mu)$ above $4\times 10^{-8}$,
making this signal difficult if not impossible to observe in Run~II.

To obtain this figure, we varied all the parameters of the GMSB model,
including the number of messengers, which we allowed to take values of
$n_5=1$, 3, and 5. We also allowed the messenger scale, $M$, to go as high
as $10^{14}\gev$. At such large $M$, one would expect to
maximize your signal since both $A$-terms and squark mixing benefit
from the running between $M$ and $m_Z$. And indeed the maximum signals
are occurring for large messenger scales. However even with such a
large scale, there is little signal in this channel compared to mSUGRA
or AMSB models. Thus a signal at the Tevatron tells us something
significant that has nothing to do with $\tan\beta$ --- it 
would be evidence {\it against}\/ gauge mediation itself~\cite{models,kl}!

\section{Conclusions}

Over the next several years, the Tevatron's physics program has a
strong opportunity to discover new physics in the rare decay
$B_s\to\mu\mu$. And if either CDF or D\O\ do observe this signature,
they will have simultaneously provided very strong evidence for a
supersymmetric world just beyond our current reach. And they will have
provided an important insight into SUSY-breaking and its mediation
mechanism. And they will have placed a lower bound on the parameter
$\tan\beta$, vitally important to future studies of SUSY. It is the
existence of this lower bound, and its value, that we have derived in
this work.

The SUSY parameter $\tan\beta$ is one of the most important
quantities which needs to be measured in a supersymmetric
world. In a top-down approach, the calculation of $\tan\beta$ from
first principles would be an important test of some underlying theory.
In a bottom-up approach, $\tan\beta$ is a necessary input to the
Yukawa sector and thus relevant to everything from Yukawa unification
to radiative electroweak symmetry breaking.
Almost all predictions that can test the form of the
underlying theory depend on $\tan\beta$ in some way, so it must be
measured as a first step in studying and testing models of SUSY. Yet
$\tan\beta$ is infamously difficult to measure, particularly 
at hadron colliders or in
rare decays --- all methods proposed to date are model and/or parameter
dependent.

The decay $B_s\to\mu\mu$ may improve the situation dramatically
because it is unusually sensitive to $\tan\beta$,
providing an opportunity to bound $\tan\beta$ if a signal is seen. 
In this
paper we showed that some very general theoretical assumptions that do
not depend on specific models allow us to put significant lower bounds
on $\tan\beta$ given a signal. In the general case with large flavor
changing effects allowed ($\delta_{LL}^{23}\sim1$), the bound is given by
Eq.~(\ref{eq:nmfvmax}), leading
to $\tan\beta\gsim 4$. This bound increases quickly with decreasing
flavor violation: for $\delta_{LL}^{23}\simeq 0.1$ the bound is pushed to
$\tan\beta\gsim 7$~(Eq.~(\ref{eq:nmfvmin})). In more typical models
in which $\delta_{LL}^{23}\lsim V_{ts}$ or $V_{ub}$, then
$\tan\beta\gsim 9$ if a signal is seen at the Tevatron.
However, the decay $B_s\to\mu\mu$ does not require new sources of
flavor changing to be present; it can be induced simply from the
flavor changing already present in the standard model CKM matrix.
In such a case a stronger bound
$\tan\beta \gsim 11$ is obtained (Eq.~(\ref{eq:mfv})). These bounds are for
a $B(B_s\to\mu\mu)>10^{-7}$; we have also shown how to scale the
bounds as a function of the branching fraction. Finally, we find that
the bounds are significantly stronger in often studied models such as
minimal supergravity.

For many purposes, a lower limit is as good as a measurement since
many observables ($m_h$, $g_\mu-2$, dark matter signals, etc) saturate
as $\tan\beta$ increases. There is also an upper limit on $\tan\beta$
of order 60 from the requirement that the theory remain perturbative
up to high (unificiation) scales. Thus a lower limit often is
tantamount to a measurement.

\section*{Acknowledgements}
We are grateful to K.~Hidaka for helpful comments on the manuscript.
This work was supported in part by 
the National Science Foundation under grant PHY00-98791 and 
by the U.S.~Department of Energy under grant DE-FG02-95ER40896.

\end{document}